\documentclass[aip,pop,amsmath,reprint]{revtex4-1}
\usepackage{amsmath} 
\usepackage{amsfonts}
\usepackage[dvips]{graphicx}
\usepackage[usenames,dvipsnames]{color}

\setkeys{Gin}{width=0.9\linewidth,keepaspectratio}

\newcommand{\eqr}[1]{Eq.\thinspace(#1)}
\newcommand{\pfrac}[2]{\frac{\partial #1}{\partial #2}}
\newcommand{\pfraca}[1]{\frac{\partial}{\partial #1}}

\newcommand{\mvec}[1]{\mathbf{#1}}

\newcommand{\refcite}[1]{Ref.\thinspace\onlinecite{#1}}

\makeatletter
\let\start@align@nopar\start@align
\let\start@gather@nopar\start@gather
\let\start@multline@nopar\start@multline
\long\def\start@align{\par\start@align@nopar}
\long\def\start@gather{\par\start@gather@nopar}
\long\def\start@multline{\par\start@multline@nopar}
\makeatother

\begin{document}

\title{Comparison of multi-fluid moment models with Particle-in-Cell
  simulations of collisionless magnetic reconnection}%
\author{Liang Wang}%
\email{lag69@unh.edu}%
\affiliation{%
  Space Science Center and Physics Department, University of New Hamsphire, Durham NH 03824, USA}%
\author{Ammar H. Hakim}%
\author{A. Bhattacharjee}%
\affiliation{%
  Center for Heliophysics, Princeton Plasma Physics Laboratory, Princeton NJ 08543-0451, USA}%
\author{K. Germaschewski}%
\affiliation{%
  Space Science Center and Physics Department, University of New Hamsphire, Durham NH 03824, USA}%

\begin{abstract}
  We introduce an extensible multi-fluid moment model in the context
  of collisionless magnetic reconnection. This model evolves full
  Maxwell equations, and simultaneously moments of the Vlasov-Maxwell
  equation for each species in the plasma. Effects like electron
  inertia and pressure gradient are self-consistently embedded in the
  resulting multi-fluid moment equations, without the need to
  explicitly solving a generalized Ohms's law. Two limits of the
  multi-fluid moment model are discussed, namely, the five-moment
  limit that evolves a scalar pressures for each species, and the
  ten-moment limit that evolves the full anisotropic, non-gyrotropic
  pressure tensor for each species. We first
  demonstrate, analytically and numerically, that the five-moment
  model reduces to the widely used Hall Magnetohydrodynamics (Hall
  MHD) model under the assumptions of vanishing electron inertia,
  infinite speed of light, and quasi-neutrality. Then, we compare
  ten-moment and fully kinetic Particle-In-Cell (PIC) simulations of a
  large scale Harris sheet reconnection problem, where the ten-moment
  equations are closed with a local linear collisionless approximation
  for the heat flux. The ten-moment simulation gives reasonable
  agreement with the PIC results regarding the structures and
  magnitudes of the electron flows, the polarities and magnitudes of
  elements of the electron pressure tensor, and the decomposition of
  the generalized Ohm’s law. Possible ways to improve the simple local
  closure towards a nonlocal fully three-dimensional closure are also 
  discussed.
\end{abstract}

\maketitle

\section{Introduction}
\label{sec:intro}

Magnetic reconnection is the process by which magnetic field line topology
changes. It is often accompanied by fast, explosive release of
magnetic energy and restructuring of macroscopic quantities of the
plasma, such as flows and thermal energy. Reconnection is widely
thought to play a critical role in many physical phenomena in
laboratory, solar, terrestrial, and astrophysical
plasmas\cite{Yamada:2010im,zweibel2009review}.

A key constraint in ideal magnetohydrodynamics (MHD) is the
frozen-flux constraint $\mathbf{E}+\mathbf{u}\times\mathbf{B}=0$, or,
equivalently $E_\parallel=0$. With this, field line topology cannot
change, and hence reconnection cannot occur. Hence, one must look
beyond the ideal MHD model to study reconnection, in particular the
generalized Ohm's law
\begin{align}
 \mathbf{E} + &\mathbf{u}\times\mathbf{B} =
 \eta\mathbf{J}
 +
 \frac{\mathbf{J}\times\mathbf{B}}{n\left|e\right|}
 -
 \frac{\nabla\cdot\mathbf{P}_{e}}{n\left|e\right|}
 \notag \\
 &
 +\frac{m_{e}}{n\left|e\right|^{2}}
    \left[\frac{\partial\mathbf{J}}{\partial t} +\nabla\cdot
    \left(\mathbf{u}\mathbf{J}
      +\mathbf{J}\mathbf{u}-\frac{\mathbf{JJ}}{n\left|e\right|}
   \right)
 \right].\label{eq:ohm}
\end{align}
Here, $\eta\mathbf{J}$ is the resistive dissipation term,
$\mathbf{J}\times\mathbf{B}/n\left|e\right|$ is the Hall term (which
by itself does not change field line topology),
$\nabla\cdot\mathbf{P}_{e}/n\left|e\right|$ is electron pressure
gradient term, and the last term is proportional to electron inertia. 
Historically, Sweet\cite{sweet} and Parker\cite{parker} proposed the first
self-consistent model for reconnection by emphasizing the role of
$\eta\mathbf{J}$, due to particle collisions, in breaking field
lines and dissipating magnetic energy. However, though confirmed by
experiments in certain regimes\cite{ji1999pp,Trintchouk2003pp}, the Sweet-Parker model generally
predicts reconnection rates that are much slower than observed in many
situations. This has prompted investigations of the role of other terms 
on the right of Eq.\thinspace(\ref{eq:ohm}) in facilitating fast reconnection.
In collisionless plasma configurations\cite{Oieroset2001nat,vtf}, the resistive 
term actually vanishes, and field-line breaking must occur due to other terms in the generalized Ohm's law.

The role of the various terms in Eq.\thinspace(\ref{eq:ohm}) for
collisionless reconnection have been extensively studied, both
analytically and with fluid and kinetic simulations\cite{Vasyliunas1975review,birn_2001a,Ma1998grl,Wang2000jgr,Cai1997pp,Kuznetsova2001jgr,Kuznetsova1998jgr}. In particular,
fully kinetic simulations indicate the importance of kinetic effects
in collisionless fast reconnection, where particle distribution function
departs from a Maxwellian near the X-point. Unfortunately, fully
kinetic simulations using Particle-in-Cell (PIC) algorithms, even with
modern high peformance codes\cite{Bowers:2008bu,psc,Lapenta2012jcp} running on the
fastest supercomputers, are extremely expensive, and generally 
feasible only in small domains (on the order of a hundred ion inertial
length boxes in in three dimensions (3D)). For global modeling of large-scale collisionless
systems like the Earth's magnetosphere, one inevitably turns to
simplified, asymptotic approaches like resistive or Hall-MHD and/or
hybrid (ion particles plus massless electron fluid) models\cite{openggcm,umichglobal,Lipatov2002book}.
Nevertheless, although useful, these simplified models are not ideal
to describe collisionless systems because they fail to include the
crucial electron kinetic physics contained in pressure tensor,
$\mathbf{P}_{e}$, by restricting it to either a scalar or ignoring it
completely (cold electron model).

Instead of expensive kinetic and/or hybrid simulations, several very
different approaches were also attempted by incorporating some kinetic
effects into simplified models. For example, Hesse \textit{et al}.\cite{HesseWinske1993grl,HesseWinske1994jgr,HesseWinske1995jgr,Kuznetsova1998jgr,Gary2000jgr,Kuznetsova2000,Kuznetsova2001jgr}
modified their hybrid code to evolve the full electron pressure tensor
supplemented by a relaxation term. Yin \textit{et al}.\cite{YinWinske2001jgr,YinWinske2002pp,YinWinske2003pp} later adopted
a similar procedure in their Hall-MHD code. By retaining
$\mathbf{P}_{e}$ their models were able to recover certain features
seen in kinetic simulations. Another approach was benchmarked
by Sugiyama \textit{et al.}\cite{Sugiyama2007jcp} and Daldorff \textit{et al.}\cite{Daldorff2014jcp}, who treated the ideal
regions with a simplified model, and the localized, non-ideal region
with a fully kinetic model. Meanwhile,
Le \textit{et al.}\cite{Le2009prl} introduced a collisionless closure relation for the
anisotropic electron pressure $\left(p_{\parallel},p_{\perp}\right)$
due to trapping of electrons in parallel electric fields. An equation
of state (EoS) computed from this closure and used in a fluid model,
was compared to PIC simulations in the strong guide-field limit,
successfully recovering structural features that are typically
observed in fully kinetic simulations\cite{Le2009prl,Le2010pp,Ohia2012prl,Egedal2013pp}. 
However, this closure is rigorously valid when the plasma is strongly magnetized, thus it fails
for anti-parallel reconnection, for example, found in the Earth's
magnetotail.

The extension of global simulation models to include collisionless 
effects is  a grand challenge problem. First,
the inclusion of the Hall term in the generalized Ohm's law is
non-trivial and often suffers from severe time-step restrictions, in
the absence of electron inertia, due to (unphysical) quadratic
dispersive whistler waves. Implicit algorithms are often used,
however, making the implementation more complicated\cite{Toth2008jcp,Chacon2002jcp,Chacon:2012,Jardin:2012iu}. 
Second, there is a compelling need to enable
multiple species in the plasma. For example, the presence of multiple
ion species, e.g., $H^{+}$ and $O^{+}$, might significantly modify the
dynamics of ring currents during a magnetospheric substorm event\cite{Daglis1999rev,Moore2001rev}.

In this paper, we use an extensible multi-fluid moment
model\cite{sumlak_2003,Loverich:2010ea,Hakim2006,Hakim:2008br,Johnson:2013vf}
to address some of the difficulties summarized above (e.g., lack of
self-consistent non-ideal terms in fluid-based codes, time step
restriction due to dispersive modes, and needs for multi-species
capability).  This model evolves the electromagnetic field using
Maxwell equations (including displacement currents), and
simultaneously evolves truncated moments of the Vlasov equation for
each species, $s$, in the plasma. Conceptually, the multi-fluid moment
model is a more complete description of the plasma (compared to
resistive and Hall-MHD): it supports electromagnetic waves, 
permits departures from quasi-neutrality, includes the Hall-term as well electron
inertia. The multi-fluid moment model is extensible in two
senses. First, it can easily handle any number of species.  Such
multi-species capability suits well the needs of modern global
models. Second, the multi-fluid moments may be easily extended to
self-consistently evolve higher order mean quantities like the
pressure tensor $\mathbf{P}_{s}$, and even the heat flux tensor,
$\mathbf{Q}_{s}$\cite{Johnson:2013vb}. In the five-moment limit, the
truncated moment equations evolve the number density, $n_{s}$,
velocity, $\mathbf{u}_{s}$, and scalar pressure, $p_{s}$. In the
ten-moment limit, ten terms are solved, replacing $p_{s}$ with the six
independent components of the pressure tensor, $\mathbf{P}_{s}$, with an
appropriate heat flux closure to truncate the equations. The
capability of the ten-moment model to self-consistently calculate the
pressure tensor in a fluid-based description is a critical step to
fill the gap between existing fluid-based global models and the
intrinsic requirement for kinetic effects to correctly model
collisionless systems. Nevertheless, it should also be emphasized that
though the incorporation of $\mathbf{P}_{s}$ itself is
straightforward, steps to find an appropriate heat flux closure are
not obvious. We will employ a simplified local approximation to a
Hammett-Perkins\cite{Hammett:1990ts} collisionless closure, and study
its role in collisionless reconnection.

On the other hand the moment equations have embedded in them the Hall 
term and electron inertia effects though the momentum equations for 
charged particle species, and no special treatment is necessary 
for a generalized Ohm's law. The symmetric form of multi-fluid
moment equations also facilitate the implementation of a locally
implicit algorithm that eliminates the time step constraints from
plasma frequency and quadratic dispersive modes, which greatly speeds
up simulation performance.

\section{The five- and ten-moment models}
\label{sec:10-5-m}

Each species in a multi-component collisionless plasma is described by
the Vlasov equation for the temporal evolution of the particle
distribution function, $f(\mvec{x},\mvec{v},t)$, defined such that
$f(\mvec{x},\mvec{v},t)d\mvec{x}d\mvec{v}$ is the number of particles
contained in a phase-space volume element $d\mvec{x}d\mvec{v}$. The
Vlasov equation may be written as
\begin{align}
  \pfrac{f}{t}+v_j\pfrac{f}{x_j} +
  \frac{q}{m}(E_j+\epsilon_{kmj}v_kB_m)\pfrac{f}{v_j} =
  0.\label{eq:boltz}
\end{align}
Here $\mvec{E}$ is the electric field, $\mvec{B}$ is the magnetic flux
density, $q$ and $m$ are the charge and mass of the plasma species and
$\epsilon_{kmj}$ is the completely anti-symmetric pseudo-tensor which
is defined to be $\pm1$ for even/odd permutations of $(1,2,3)$ and
zero otherwise. Summation over repeated indices is assumed. The moment
equations shown below are derived for each species independently, so
the species indices are dropped. The electromagnetic fields are
evolved using Maxwell equations
\begin{align}
  \nabla\times \mvec{E} &= -\pfrac{\mvec{B}}{t}, \label{eq:mecurl} \\
  \nabla\times \mvec{B} &=
  \mu_0\mvec{J}+\frac{1}{c^2}\pfrac{\mvec{E}}{t}. \label{eq:mbcurl}
\end{align}
Here $\mu_0$ and $\varepsilon_0$ are the permeability and permittivity
of free space, $c=(\mu_0\varepsilon_0)^{-1/2}$ is the speed of light
and $\mvec{J}$ is the current density defined by
\begin{align}
  \mvec{J} &\equiv \sum qn\mvec{u}. \label{eq:current}
\end{align}
The summations in Eq.\thinspace(\ref{eq:current}) are over all species
present in the plasma. The number density $n(\mvec{x},t)$ and mean
velocity $\mvec{u}(\mvec{x},t)$ are defined by
\begin{align}
  n &\equiv \int f d\mvec{v}, \label{eq:numberdensity} \\
  u_j &\equiv \frac{1}{n} \int v_j f d\mvec{v}, \label{eq:velocity}
\end{align}
where $d\mvec{v}=dv_1dv_2dv_3$ represents a volume element in velocity
space. 

The moment equations are obtained in the usual way by multiplying the Vlasov
equation by “moments” of the velocities and integrating over velocity space. For example, in
addition to the number density and mean velocities defined by
Eqs.\thinspace(\ref{eq:numberdensity}) and (\ref{eq:velocity}) the
following higher order moments are defined:
\begin{align}
  \mathcal{P}_{ij} &\equiv m \int v_i v_j f d\mvec{v}, \label{eq:mcp} \\
  \mathcal{Q}_{ijk} &\equiv m \int v_i v_j v_k f d\mvec{v}.
  \label{eq:mcq}
\end{align}
With these one obtains the set of \emph{exact} moment equations
\begin{align}
  &\pfrac{n}{t}+\pfraca{x_j}(nu_j) = 0, \label{eq:num} \\
  &m\pfraca{t}(nu_i) + \pfrac{\mathcal{P}_{ij}}{x_j}
  =
  nq(E_i+\epsilon_{ijk}u_jB_k), \label{eq:mom} \\
  &\pfrac{\mathcal{P}_{ij}}{t} + \pfrac{\mathcal{Q}_{ijk}}{x_k}
  =
  nqu_{[i}E_{j]}
  + \frac{q}{m}\epsilon_{[ikl}\mathcal{P}_{kj]}B_l. \label{eq:press}
\end{align}
In these equations square brackets around indices represent the
minimal sum over permutations of free indices needed to yield
completely symmetric tensors. For example $u_{[i}E_{j]} = u_iE_j +
u_jE_i$.  Equations\thinspace(\ref{eq:num})--(\ref{eq:press}) are 10
equations (1+3+6) for 20 unknowns ($\mathcal{Q}_{ijk}$ has 10
independent components). In general any finite set of exact moment
equations will always contain more unknowns than equations. Writing
\begin{align}
  \mathcal{Q}_{ijk} &= {Q}_{ijk} + u_{[i}\mathcal{P}_{jk]} - 2nm u_i u_j u_k \
  \label{eq:heat_ijk}
\end{align}
to close this system of equations we need a closure approximation for
the heat-flux, ${Q}_{ijk}$, defined as
\begin{align}
  {Q}_{ijk} &\equiv m \int (v_i-u_i) (v_j-u_j)(v_k-u_k) f d\mvec{v}.
\end{align}
The system of equations
Eqns.\thinspace(\ref{eq:num})--(\ref{eq:press}), closed with an
approximation for the divergence of the heat-flux tensor (along with
Maxwell equations), are the \emph{ten-moment
  equations}\cite{Buneman1961,Hakim:2008br}. For a plasma of $S$
species, they consist of $10S+8$ equations.

In the ten-moment model, the pressure tensor can have arbitrary
anisotropy and orientation. The ``double adiabatic'' law of
Chew-Goldberg-Low\cite{cgl-1956} (CGL), on the other hand, describes
the pressure evolution of the principal components ($p_\parallel,
p_\perp$) in a strong magnetic field, and is contained the pressure
evolution equation, \eqr{\ref{eq:press}}. In general, the principal
axes of the pressure tensor need \emph{not} align with the magnetic
field, and this is consistently handled by the ten-moment model.

Before describing our closure approximation for the heat-flux, we look
at the \emph{five-moment} limit of the ten-moment equations. Writing
$\mathcal{P}_{ij} = {P}_{ij} + nm u_i u_j$ where
\begin{align}
  {P}_{ij} \equiv m \int (v_i-u_i) (v_j-u_j) f d\mvec{v},
\end{align}
we get that $\mathcal{E}\equiv \mathcal{P}_{ii}/2$ is the total fluid
(thermal plus kinetic) energy
\begin{align}
  \mathcal{E}\equiv \frac{1}{2}\mathcal{P}_{ii} = \frac{3}{2}p + \frac{1}{2}mn \mvec{u}^2,
\end{align}
where $p\equiv P_{ii}/3$ is the fluid scalar pressure. Hence, taking
(half) the trace of \eqr{\ref{eq:press}} gives the evolution
equation for the total fluid energy
\begin{align}
  \pfrac{\mathcal{E}}{t} + \frac{1}{2}\pfrac{\mathcal{Q}_{iik}}{x_k} 
     = nq\mvec{u}\cdot\mvec{E}, \label{eq:ex-fl-energy}
\end{align}
where, from \eqr{\thinspace\ref{eq:heat_ijk}} we get
\begin{align}
  \frac{1}{2}\mathcal{Q}_{iik} = q_k + u_k(p + \mathcal{E}) +
  u_i\pi_{ik}, \label{eq:q-trace}
\end{align}
where $q_k \equiv Q_{iik}/2$ is the \emph{heat-flux} vector, and
$\pi_{ij}=P_{ij}-p\delta_{ij}$, is the viscous stress tensor. 

The energy equation, \eqr{\ref{eq:ex-fl-energy}}, along with
\eqr{\ref{eq:q-trace}} is exact. The \emph{ideal} five-moment model is
formally obtained by the closure approximation $q_k=0$ and
$\pi_{ij}=0$. That is, in the five-moment model, instead of the
pressure tensor equation, \eqr{\ref{eq:press}}, we use the scalar
equation for the total fluid energy
\begin{align}
  \pfrac{\mathcal{E}}{t} + \pfraca{x_k}\left(u_k(p+\mathcal{E})\right)
     = nq\mvec{u}\cdot\mvec{E}, \label{eq:fl-energy}
\end{align}
With these approximations the system of equations for number density,
momentum density and energy are closed, and consist of five equations
for each species of the plasma, in addition to Maxwell equations. For
a plasma with $S$ species, we have $5S+8$ equations. Note that for an
electron-ion plasma, this model still retains the electron inertia
term as well as separate (but adiabatic) pressure equations for each
species. Hence, in the low-frequency limit (ignoring plasma
oscillations and electromagnetic waves) the five-moment model is a
generalization of the Hall-MHD model. One can obtain the Hall-MHD
equations \emph{formally} by taking the limit $m_e\rightarrow 0$,
$\epsilon_0\rightarrow 0$ (infinite speed of light), which also
implies quasi-neutrality, $n_e=n_i$. Analytical and numerical
comparisons of the physics of the five-moment and Hall-MHD models,
including for reconnection, was made by Srinivasan and
Shumlak\cite{Srinivasan:2011gd}.

We should point out that for both the five- and ten-moment model the
full (including displacement currents) Maxwell equations are solved. A
generalized Ohm's law is not used to evolve the electric field. This
makes it simpler to incorporate inertia effects as well as pressure
gradient terms. However, for explicit schemes, the speed of light
needs to be resolved. For the problems studied here, this is not a
severe constraint, but can be relaxed using implicit methods for the
fields.

\section{\label{sec:closure}Towards a collisionless heat-flux closure}

To close the ten-moment equations, a closure relation is needed for
the divergence of the heat-flux tensor. In space plasma applications
collisions are negligible, and hence closures based on expansion in
small mean-free-path (i.e., Chapman-Enskog expansion) are not
appropriate. However, as is well known, developing closures applicable
to collisionless systems is difficult. In fact, for problems dominated
by kinetic effects, for example, in which the physics relies on
detailed structure of the distribution function in velocity space,
recourse must be made to solving the Vlasov-Maxwell equations directly
using a PIC or a continuum code.

One approach to collisionless closures is to design closures that
reproduce the exact kinetic results for linear problems. There have 
been two major efforts. The first\cite{Chang:1992hk},
involves determining the \emph{exact} solution to the linearized
Vlasov-Maxwell equation, and then taking the appropriate integrals
over velocity space to find an expression for the unclosed fluid
moment. This leads to, in Fourier space, expressions that are
complicated functions of the plasma response function,
$R(\zeta)=1+\zeta Z(\zeta)$, where $Z(\zeta)=\pi^{-1}\int dt
\exp(-t^2)/(t-\zeta)$ is the usual plasma dispersion function. In
physical space these expressions are equivalent to a non-local
closure, involving integration along field lines or all space. In the
second approach, instead of seeking an exact solution to the
Vlasov-Maxwell equations, one uses an \emph{approximation} based on a
$n$-pole Pade-series representation of the response function. This
technique was pioneered by Hammett and Perkins\cite{Hammett:1990ts},
and since then has been used in several applications, specially for
the moments of the gyrokinetic equations, i.e. to develop
\emph{gyrofluid} models\cite{Hammett:1992tf,Snyder:1997fs}. The
trapped gyro-Landau-fluid model\cite{Kinsey:2008gb}, based on this
work, for example, is at present the leading reduced model for
turbulent fluxes used in transport time-scale simulations of tokamak
core plasmas. Note that the Hammett-Perkins type closure also leads to
non-local closures, requiring integrations along field lines. Goswami
and co-workers\cite{Goswami:2005cl} have extended this Pade-series
technique to full 3D fluids (i.e., not gyrofluid) in the presence of a
strong magnetic field. Physical space implementation of their
closures, however, have not been extensively used (as far as we know)
in numerical simulations. An overview of these, and CGL-like,
collisionless closures is provided by Chust and
Belmont\cite{Chust:2006iq}.

As a first step towards a non-local closure, we have implemented a
simple \emph{local} approximation that captures some of the physics of
collisionless damping. In addition to being local, we also ignore the
preferred direction imposed by the magnetic field in the closure. One
should note that even though this approximation is rather crude, the
full non-linear ten-moment equations are solved, with the closure only
affecting a single term (divergence of heat-flux) in the pressure
tensor equation. As shown below, even this simple closure gives
reasonable agreement with PIC results for Harris sheet reconnection,
indicating that better collisionless closures (left to future work)
will further improve the agreement, at least for certain collisionless
problems.

In the Hammett-Perkins approach the following form of the perturbed
heat-flux (in one dimension) leads to a three-pole Pade-series
approximation to the plasma response function.
\begin{align}
  \tilde{q}(k) = -n_0\chi_1 \frac{2^{1/2}}{|k|}i k v_t \tilde{T}(k),
  \label{eq:hp-q}
\end{align}
where, tildes indicated perturbations around equilibrium quantities and
$k$ is the wave number. Further, $n_0$ is the equilibrium number
density, $v_t=\sqrt{T/m}$ is the thermal velocity, and $\chi_1$ is
some constant. In physical space, performing the inverse Fourier
transform leads to an integral relation for the heat-flux (see
Eq.\thinspace(8) in\cite{Hammett:1990ts}), which is related to the
Hilbert transform of the temperature. Several methods may be
constructed to rapidly perform this integration, which, in a
magnetized plasma, is an integration along field lines. One may, for
example, assume that the temperature dependence along a field line can
be expressed as an expansion, each term of which has an easily
computed Hilbert transform\cite{James:2007vk}. Another possibility is
to expand the sign function ($k/|k|$) in an easily invertible
series\cite{dimits2012}. In a magnetized plasma, this leads to a few
one-dimensional tridiagonal matrix inversions, which may be computed
rapidly. Another approach is to evaluate the Hilbert transform
using fast multipole methods\cite{Greengard1987}.

Here, instead we make a further approximation by simply using a local
approximation of the heat-flux, obtained by picking a typical
wave-number, $k_0$. As we are only interested in the divergence of the
heat-flux, taking derivative of \eqr{\ref{eq:hp-q}} and generalizing to
the three-dimensional unmagnetized plasmas (i.e. without any preferred
direction). Upon dropping order-unity constants, we have
\begin{align}
  ik_m Q_{ijm}(k) = v_t |k| \tilde{T}_{ij}(k) n_0,
\end{align}
where, now, $k=|\mvec{k}|$, $\tilde{T}_{ij}(k) = (\tilde{P}_{ij}(k) -
T_0\tilde{n}\delta_{ij})/n_o$ is the perturbed temperature (note that $k$ is the
wave-number and not a tensor index). This can be written as
\begin{align}
  ik_m Q_{ijm}(k) = v_t |k| ({P}_{ij}(k) - \bar{P}_{ij} - \delta_{ij}(n(k) - \bar{n})\bar{T}),
\end{align}
where bars on quantities denote ``equilibrium'' quantities around which
linearization has been performed. Note that at this point this
expression is simply a generalization of
\eqr{\ref{eq:hp-q}} to unmagnetized plasma. As such, using this in the linearized ten-moment
equation will yield a \emph{three}-pole Pade approximation to the
plasma response function.

This closure, in physical space, still needs an integration over all
space to compute. To convert it into a local approximation, we replace
the continuous wave-number, $k$ with a typical wave-number, $k_0$,
which defines a scale over which collisionless damping is thought to
occur. Hence, in physical space we can write
\begin{align}
  \partial_m Q_{ijm} \approx v_t |k_0| ({P}_{ij} -
  \bar{P}_{ij} - \delta_{ij}(n - \bar{n})\bar{T}).  
\end{align}
Note that the equilibrium quantities (represented by overbars) must be
evaluated using an averaging over a box of typical size
$1/k_0$. We can make the further
approximation that $\bar{P}_{ij}= p\delta_{ij}$ and $\bar{n}=n$. This
finally leads to the closure used here,
\begin{align}
  \partial_m Q_{ijm} \approx v_t |k_0| ({P}_{ij} - p\delta_{ij}).
  \label{eq:local-q-closure}
\end{align}

Use of \eqr{\ref{eq:local-q-closure}} in the ten-moment equations
appears, superficially, like collisional relaxation of the pressure
tensor, with the collision frequency $\nu=|k|v_t$. Such collisional
relaxation terms have been used before, for example, by Hesse \textit{et al}.\cite{HesseWinske1993grl,HesseWinske1994jgr,HesseWinske1995jgr,Kuznetsova1998jgr,Gary2000jgr,Kuznetsova2000,Kuznetsova2001jgr}, Yin \textit{et al}.\cite{YinWinske2001jgr,YinWinske2002pp,YinWinske2003pp}, Hakim\cite{Hakim:2008br}, Johnson\cite{Johnson:2013vf} and
Brackbill\cite{Brackbill:2011if}. However, the reconnection problems
we are interested in are collisionless. Hence, the selection of a
``collision frequency'' seems arbitrary. Also, the more accurate
Hammett-Perkins type closures are \emph{non-local} and need not relax
all components of the pressure tensor, but will involve non-local
temperature gradient drives, with damping occurring at all
wave-numbers. Further, extending the moment system to the next larger
set, the \emph{twenty-moment} model (in which all ten independent
components of $Q_{ijk}$ are evolved), combined with a similar
Hammett-Perkins type closure (see unnumbered equation after
Eq.\thinspace(10) in\cite{Hammett:1990ts}) will lead to a four-pole
Pade approximation of the plasma response function, with \emph{no}
``collisional'' damping in the pressure equation. If, on the other
hand, a collision operator is used, there will appear a relaxation
term also on the right hand side of the pressure equation, in addition
to inter-species momentum drag.



We should emphasize that the approximate local form,
\eqr{\ref{eq:local-q-closure}}, is used here to illustrate the utility
of ideas from collisionless closure theory in multi-fluid moment
equations. All other terms in the ten-moment equations are retained
for each species of the plasma. Incorporation of more sophisticated
non-local closures is part of future work.



\section{Overview of numerical methods}

To solve the system of five- and ten-moment equations, we use a second
order, well-centered, locally implicit scheme. This scheme is
described in\cite{Hakim2014,Loverich:2013cr} and not described here in
detail. Using a locally implicit, operator splitting approach, the
plasma-frequency and Debye length scales can be eliminated, leading to
significant speedup in multi-fluid simulations, specially with
realistic electron/ion mass ratios, even when using an explicit
scheme. The speed of light constraint still exists, however, can be
greatly relaxed, using artificially low values for the speed of light
and/or sub-cycling Maxwell equations. Of course, an implicit Maxwell
solver, or a reduced set of electromagnetic equations like the Darwin
approximation\cite{birdsallbook}, can also relax the time-step
restrictions. In either case, though, a fully implicit approach is
needed, and are not considered in this paper.

For the hyperbolic homogenous part of the equations, we use a
dimenionsally-split finite-volume (FV) wave-propagation
scheme\cite{leveque_book_2002, Hakim2006}. This scheme is based on
solving the Riemann problem at each interface and using this to
compute numercial fluxes, which are then used to construct a
second-order scheme. Although we use the full eigensystem of both the
five- and ten-moment equations, we do not use Roe
averages\cite{roe_1981}, although they can be computed for the
ten-moment equations\cite{Brown:1996wy}. To ensure that the number
density and (diagonal components of) pressure remain positive, on
detection of a negative density/pressure state, we retake the
homogeneous step using a diffusive, but positive,
Lax-flux\cite{bouchut2004}. Although Lax-flux adds diffusion, the
scheme still conserves particles and energy. Spatial accuracy is not
severely affected as these positivity violations only occur
infrequently, or not at all. The usual ``trick'' of setting a
density/pressure ``floor'', or adding a localized diffusion term, is
not recommended as it introduces uncontrolled particle and energy
conservation errors in the solution.
\section{\label{sec:5m-hall}Five-Moment Model in the Hall MHD Limit}

In this section, we first start from an analytical point of view by
introducing the limits under which the five-moment equations formally
reduce to the Hall MHD equations, and briefly discuss the consequent
differences between the two models. Then, we present five-moment simulations
using different electron masses, $m_{e}$, to demonstrate the reduction
under the limit, $m_{e}\to0$. The results will be compared with a
previous $m_{e}$ scanning study using a Hall MHD code. This analytical
and numerical study clarifies the relation between the five-moment
and Hall MHD models, and reveals the underlying physics content of the former.

The Hall MHD equations can be formally obtained from the five-moment
equations by taking the limit $\varepsilon_{0}\to0$ and $m_{e}\to0$.
Assuming $\varepsilon_{0}\to0$ essentially indicates infinite speed
of light ($1/c^{2}\to0$) and quasi-neutrality ($n_{e}=n_{i}$). Consequently,
the displacement current in Eq.(\ref{eq:mbcurl}) can be neglected
to yield
\begin{equation}
\mu_{0}\mathbf{J}=\nabla\times\mathbf{B},\label{eq:hall-E}
\end{equation}
while the continuity equations, Eq.(\ref{eq:numberdensity}), for
electrons and ions, reduce to a single one,
\begin{equation}
\frac{\partial\rho}{\partial t}+\nabla\cdot\left(\rho\mathbf{u}\right)=0,\label{eq:hall-density}
\end{equation}
and the momentum equations, Eq.(\ref{eq:velocity}, replacing $\partial\mathcal{P}_{ij}/\partial x_{j}$
by $\partial\left(p+nmu_{i}u_{j}\right)/\partial x_{j}$), for electrons
and ions can be added to get
\begin{equation}
\frac{\partial\left(\rho\mathbf{u}\right)}{\partial t}+\nabla\cdot\left(\rho\mathbf{u}\mathbf{u}\right)=\mathbf{J}\times\mathbf{B}-\nabla p.\label{eq:hall-momentum}
\end{equation}
Here, $\rho\equiv\left(m_{i}+m_{e}\right)n$ is the bulk plasma mass
density, $\mathbf{u}\equiv\left(m_{i}\mathbf{u}_{i}+m_{e}\mathbf{u}_{e}\right)/\left(m_{i}+m_{e}\right)$
is the bulk plasma velocity, and $p\equiv p_{e}+p_{i}$ is the bulk
plasma pressure. Finally,  neglecting terms of order $O\left(m_{e}/m_{i}\right)$,
the electron/ion momentum equations can be subtracted to get the generalized
Ohm's law, Eq.(\ref{eq:ohm}). Many implementations of Hall MHD further
omit all electron inertia effects (i.e., $m_{e}=0$), then Eq.(\ref{eq:ohm})
becomes 
\begin{equation}
\mathbf{E}+\mathbf{u}\times\mathbf{B}=\frac{\mathbf{J}\times\mathbf{B}}{n\left|e\right|}-\frac{\nabla p_{e}}{n\left|e\right|}.\label{eq:hall-ohm}
\end{equation}

Hall MHD model and five-moment model are similar in that they both
retain the Hall effects and both evolve scalar pressures. The evident
differences are that the five-moment model is a more complete description
of the plasma, while the Hall MHD model does not describe charge separations,
plasma oscillations, or electromagnetic waves. Furthermore, many implementations
of Hall MHD omit electron inertia, and require an explicit resistivity
term, $\eta\mathbf{J}$, or a hyperrestivity term, $\eta_{H}\nabla^{2}\mathbf{J}$,
on the right-hand side of Eq.(\ref{eq:hall-ohm}) to break the frozen-flux
constraint. In contrast, in the five-moment model, electron and ion
inertia are self-consistently embedded within the momentum equations,
and are responsible for breaking the frozen-flux constraint.

\begin{figure}
\includegraphics[width=0.98\columnwidth]{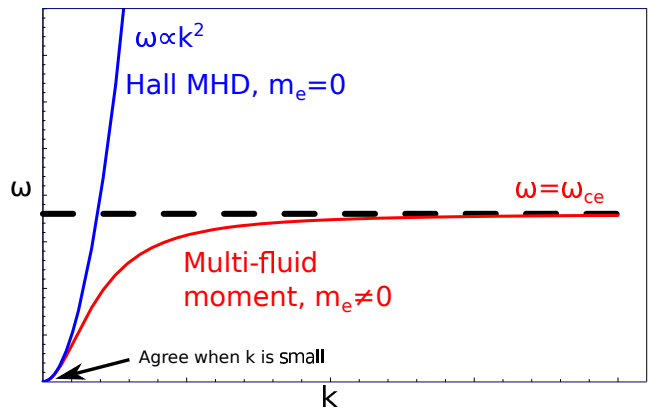}\caption{\label{fig:whistler-5m-hall}
Dispersion curves for the whistler waves of the multi-fluid plasma model with finite electron mass
  (red curve), the Hall-MHD model with zero electron mass
  (blue curve). Finite electron inertia allows the whistler wave to have an asymptote at the electron cyclotron frequency (marked by the black horizontal dashed line) whereas ignoring electron inertia causes it to grow without bound as seen in the Hall-MHD model.
}
\end{figure}

Omitting electron inertial in the Hall MHD model also leads to a whistler
wave mode with a quadratic dispersion relation, $\omega\propto k^{2}$.
The resulting whistler speed, $v_{\text{whistler}}\propto k$, grows
without bound in the short wavelength limit ($k\gg1$), severely restricting
the time step size in an explict Hall MHD simulation. This is a well-known
restriction of explicit Hall MHD simulations, and generally requires
non-trivial treatments like hyperresistivity, or complicated implicit algorithms. In comparison,
in the five-moment model, just like in a realistic cold plasma, the
whistler mode asymptotes to a resonance at electron cyclotron frequency,
$\Omega_{ce}$, in the short wavelength limit, imposing a time step
restriction bounded only by $\Omega_{ce}^{-1}$. The dispersion relation
curves of the whistler mode in the the Hall MHD model (omitting $m_{e}$)
and the five-moment model are illustrated in Fig. \ref{fig:whistler-5m-hall}.

Sullivan \textit{et al}.\cite{sullivan2009} performed a set of Hall
MHD simulations with different $m_{e}$ to study the current sheet
formation in a 2D coalescence problem, and found the scalings of the
length and thickness of the electron current layer versus $m_{e}/m_{i}$.
Here, we perform five-moment simulations of the same setup using different
$m_{e}/m_{i}$ as well to find corresponding scalings in the five-moment
model, and compare with their Hall MHD results.

The initial equilibrium is depicted in Fig.(1) of \refcite{sullivan2009},
consisting of four flux tubes sustained by four out-of-plane current
channels. A doubly periodic domain is employed, $-L/2<x,y<L/2$, where
$L=12.8d_{i0}$, and $d_{i0}=c/\omega_{pi0}$ is the ion inertia length
based on a characteristic density $n_{0}$. For convenience, the equilibrium
setup is summarized below:
\begin{eqnarray}
B_{x} & = & \left(2\pi/12.8\right)B_{x0}\sin\left(2\pi x/L\right)\cos\left(2\pi y/L\right),\label{eq:sullivan-ic-Bx}\\
B_{y} & = & \left(2\pi/12.8\right)B_{x0}\cos\left(2\pi x/L\right)\sin\left(2\pi y/L\right),\label{eq:sullivan-ic-By}\\
n & = & n_{0}\left[1+\frac{B_{x0}^{2}-B^{2}}{2\mu_{0}n_{0}k_{B}\left(T_{i}+T_{e}\right)}\right]\label{eq:sullivan-ic-n}.
\end{eqnarray}
Here, $T_{e}=T_{i}=B_{x0}^{2}/2\mu_{0}n_{0}k_{B}$, and $B_{x0}$
is chosen such that $v_{Ax0}\equiv\sqrt{B_{x0}^{2}/\mu_{0}n_{0}m_{i}}=0.1c$.
Ions are then perturbed to stream into the X-point along the anti-diagonal
direction at the following velocity:
\begin{eqnarray}
\delta\mathbf{u}_{i} & = & u_{0}\left[\sin\left(2\pi y/L\right)\hat{\mathbf{x}}+\sin\left(2\pi x/L\right)\hat{\mathbf{y}}\right],\label{eq:sullivan-ic-pert}
\end{eqnarray}
where $u_{0}=0.1v_{Ax0}$. These ion flows bring the two anti-diagonal
flux tubes towards each other, and form a current sheet at the X-point
along diagonal direction. We will test four different electron mass
ratios, $m_{e}/m_{i}=1/25$, $1/100$, $1/256$, and $1/400$. 

The typical evolution of this system is that the current layer shrinks
in both length and thickness before forming a stable structure around
$t=10\Omega_{ci}^{-1}$. The time dependences of lengths and thickneses
of the electron current layers in the Hall MHD simulations can be
found in the top panels of Fig.(11) and (13) of \refcite{sullivan2009}.
Our five-moment simulations turn out produce qualitatively very similar
time dependences. In the following study, we will focus on the later
quasi-steady stages of different five-moment runs, when lengths and
thicknesses of the electron current layers remain stable.

\begin{figure*}
\includegraphics[width=1.98\columnwidth]{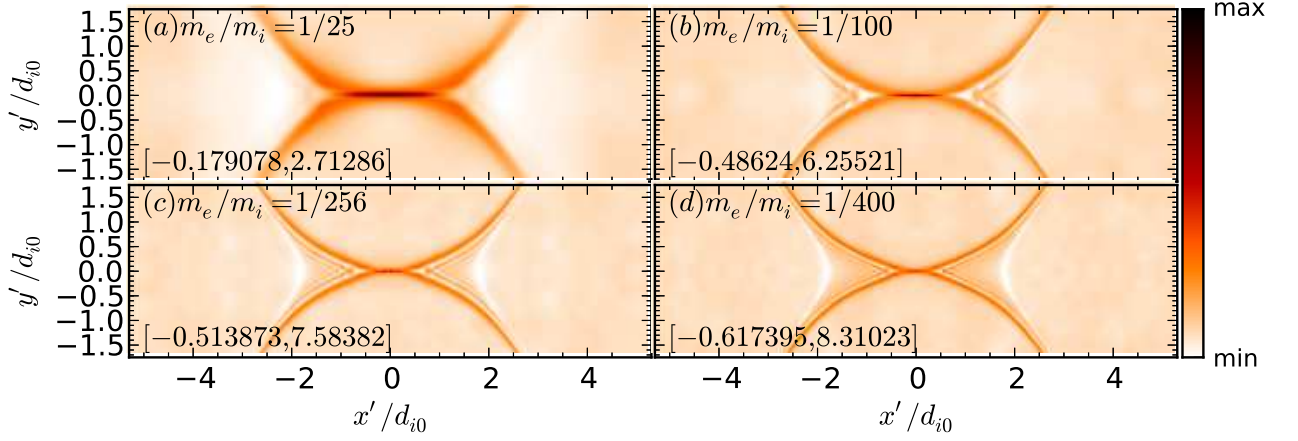}\caption{\label{fig:sullivan-Jze-visual}Out-of-plane electron current, $J_{z,e}$,
in the unit of $n_{0}\left|e\right|v_{Ax0}$ from the five-moment
simulations motivated by Sullivan \textit{et al}.\cite{sullivan2009}. Different panels correspond
to different mass ratios, $m_{e}/m_{i}$, as marked at upper left
of each panel. The values in lower right square brackets indicate
the range of $J_{z,e}$ values within the region shown.}
\end{figure*}

Fig. \ref{fig:sullivan-Jze-visual} shows the color contours of out-of-plane
electron current density, $J_{z,e}$, in the quasi-steady states from
different runs. There is a clear transition from the thick, diffusive,
and relatively more elongated current layer for the case of $m_{e}/m_{i}=1/25$
(top-left), to the short, X-like structure for the case of $m_{e}/m_{i}=1/400$
(bottom right), bridged by the intermediate pictures of $m_{e}/m_{i}=1/100$
(top right) and $1/256$ (bottom left). This transition qualitatively
agrees with the Hall MHD simulations, shown in Fig.(2 and 3) of \refcite{sullivan2009}.
The recovery of the X-like structure in the five-moment run with $m_{e}/m_{i}=1/400$
confirms that this run is approaching the limit of Hall MHD with $m_{e}=0$,
since such structure is typically seen in Hall MHD simulations with
$m_{e}=0$.

\begin{figure*}
\includegraphics[width=0.98\columnwidth]{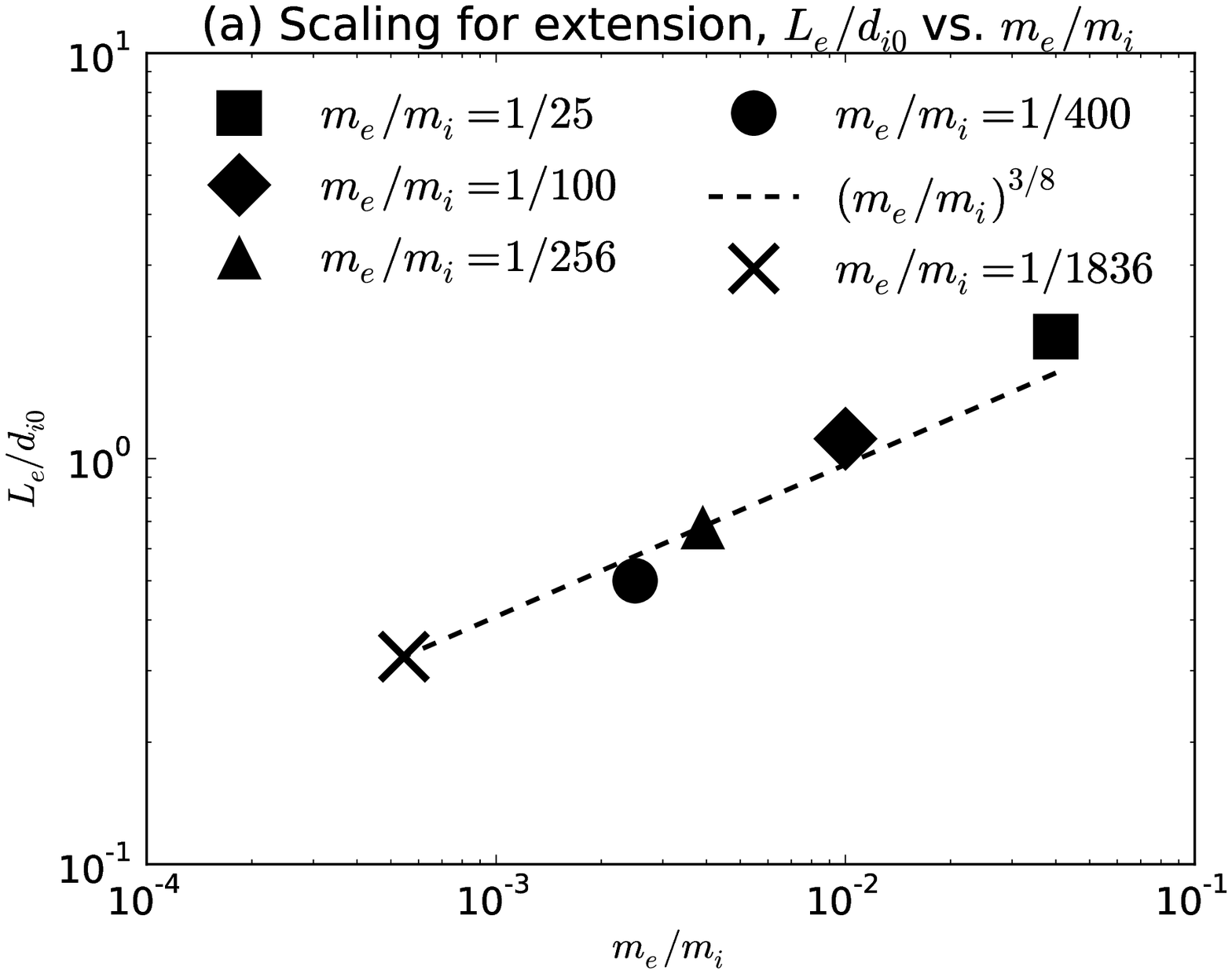}\includegraphics[width=0.98\columnwidth]{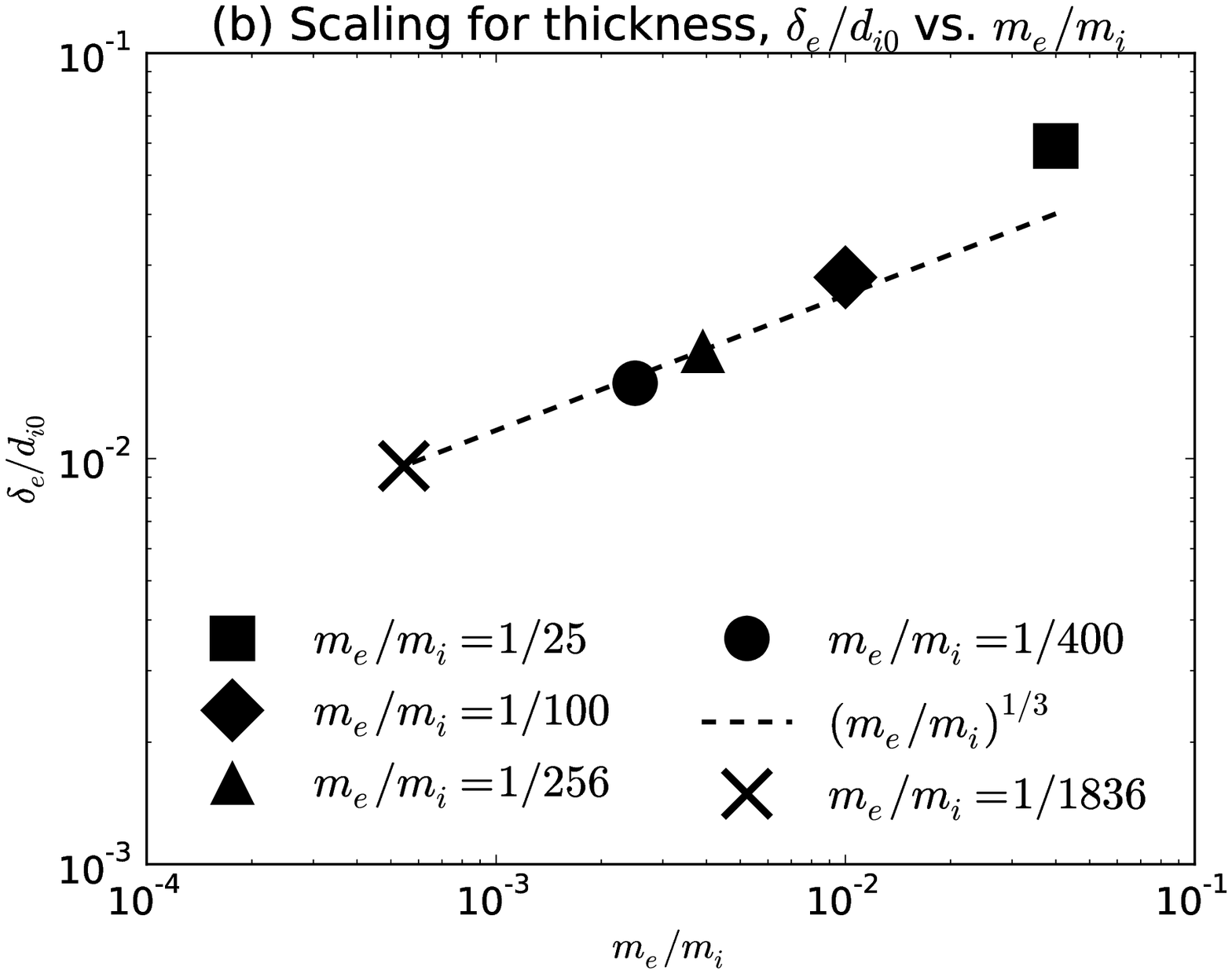}\caption{\label{fig:sullivan-scalings}Left: $L_{e}/d_{i0}$ vs. $m_{e}/m_{i}$.
The solid squares, diamonds, triangles, and circles represent five-moment
simulations with $m_{e}/m_{i}=1/25$, $1/100$, $1/256$, and $1/400$,
respectively. The dashed line indicates an estimated scaling $\sim(m_{e}/m_{i})^{3/8}$,
with the black cross on the lower left end marking an extrapolated
value of $L_{e}/d_{i0}$, for $m_{e}/m_{i}=1/1836$. Right: $\delta_{e}/d_{i0}$
vs. $m_{e}/m_{i}$. The dashed line indicates an estimated scaling
of $(m_{e}/m_{i})^{1/3}$, with the black cross marking an extrapolated
value of $\delta_{e}//d_{i0}$ for $m_{e}/m_{i}=1/1836$.}

\end{figure*}

Fig. \ref{fig:sullivan-scalings} shows the scalings of median stable
plateau values of $L_{e}$ vs $m_{e}/m_{i}$ (left panel), and median
stable plateau values of $\delta_{e}$ vs $m_{e}/m_{i}$ (right panel).
Here, the length of the current layer, $L_{e}$, is defined by the
distance between the peaks of electron outflow velocities, $u_{x,e}$,
and the thickness, $\delta_{e}$, is the full width at half maximums
of $J_{z,e}$. In both panels of Fig. \ref{fig:sullivan-scalings},
the solid squares, diamonds, triangle, and circles denote data from
runs with $m_{e}/m_{i}=1/25$, $1/100$, $1/256$, and $1/400$, respectively.
The dashed line in the left panel represents an estimated scaling
$L_{e}\propto\left(m_{e}/m_{i}\right)^{3/8}$. The cross on the lower
left end denotes an extrapolated value of $L_{e}\sim12\text{ to }17d_{e0}$,
or $0.28\text{ to }0.4d_{i0}$, for a physical mass ratio, $m_{e}/m_{i}=1/1836$.
In the right panel, the dashed line represents an estimated scaling
$\delta_{e}\propto\left(m_{e}/m_{i}\right)^{1/3}$, where the cross
denotes an extrapolated value of $\delta_{e}\sim0.4\text{ to }0.6d_{e0}$
for the physical mass ratio. Equivalent scalings from Hall MHD simulations
are shown in the bottom panels of Fig(11) and (13) of \refcite{sullivan2009}.
The extrapolated values of $L_{e}$ are in good agreement, but the
Hall MHD simulations predicted a significantly larger $\delta_{e}\sim1d_{e0}$. 

It should be clarified that, Sullivan \textit{et al}. used a hyperresistivity 
in their Hall MHD simulations to break the
frozen-flux constraint (necessary when $m_{e}=0$) and to maintain numerical stability,
while our five-moment code does not include such a term. The 
hyperresistivity can control the length of the current sheet\cite{Chacon2007,sullivan2009}. However, when Sullivan \textit{et al}.
studied the effects of electron inertia, a tiny, fixed hyperresistivity was used
in different runs, thus the influences on different runs due 
to this term should be similar and relatively small.
Meanwhile, isothermal EoS were employed in their Hall
MHD simulations, while the energy equations evolved in our five-moment
simulation essentially indicate adiabatic EoS. Taking into account
such inconsistencies between the numerical codes, and also the possible
measurement errors due to the limited number of data points and the
vast range of $m_{e}/m_{i}$ to be fitted, the overall agreements
between the five-moment and the Hall MHD simulations are remarkably
good.

\section{\label{sec:gem}2D Anti-Parallel Reconnection In A Harris Sheet}

In this section, we investigate the ten-moment limit of the multi-fluid
moment model, particularly its capability of evolving full electron
pressure tensor, $\mathbf{P}_{e}$. To this end, ten-moment simulations
are performed of 2D anti-parallel reconnection in a Harris sheet.
We examine the roles of $\mathbf{P}_{e}$, in controlling the
structures of electron current layer, and in supporting the reconnection
electric field near the X-point. For comparison, PIC and five-moment
simulations are performed with the same setup. The PIC simulation
is fully kinetic, while the five-moment simulation permits scalar pressures
only. The transition from five-moment to ten-moment and then to PIC
thus forms a complete set of comparison. The PIC simulation are performed
using the numerical code \textit{PSC}\cite{psc}.

Two ten-moment simulations are performed, with $k_{e0}=k_{i0}=1/10^{-4}d_{e0}$
and $1/d_{e0}$, respectively, where $d_{e0}$ is the electron inertia
length due to an asymptotic number density $n_{0}$. $k_{e0}$ and
$k_{i0}$ are the constant wave-numbers defining electron and ion
heat-flux approximations in the form of \eqr{\ref{eq:local-q-closure}}. The $k_{e0}=k_{i0}=1/10^{-4}d_{e0}$
run approximates the limit of $k_{e0},k_{i0}\to\infty$, approaching
the five-moment model. The $k_{e0}=k_{i0}=1/d_{e0}$ run, instead,
aims at approaching the PIC run. $k_{e0}$ and $k_{i0}$ are chosen
so as to approximately capture the characteristic length scale near
the X-point, since previous studies of similar problems indicated
that the electron current layer tends to thin down to a thickness
comparable to $d_{e0}$. Our primary task is then to find the similarities
and differences between this particular ten-moment run and the PIC
run. For convenience, we call this run the \textit{targeted}
run in the rest of this study.

\subsection{Numerical setup }

We employ a 2D simulation domain that is periodic in $x$ ($-L_{x}/2<x<L_{x}/2$)
and is bounded by conducting walls in $y$ at $y=\pm L_{y}/2$. Here
$L_{x}=100d_{i0}$, $L_{y}=50d_{i0}$, where $d_{i0}=\sqrt{m_{i}/\mu_{0}n_{0}\left|e\right|^{2}}$
is the ion inertia length due to $n_{0}$. The initial equilibrium
is a single Harris sheet where magnetic field and number densities
are specified by 
\begin{equation}
\mathbf{B}_{0}=B_{0}\tanh\left(y/\lambda_{B}\right)\hat{\mathbf{x}},
\end{equation}
and 
\begin{equation}
n_{e}=n_{i}=n_{0}\text{sech}^{2}\left(y/\lambda_{B}\right)+n_{b},
\end{equation}
respectively. The total current, $\mathbf{J}_{0}=\nabla\times\mathbf{B}_{0}/\mu_{0}$,
is decomposed according to $J_{ze0}/J_{zi0}=T_{i0}/T_{e0}$, where
the initial temperatures $T_{e0}$ and $T_{i0}$ are constant. A sinusoid
perturbation is then applied on the magnetic field according to $\delta\mathbf{B}=\hat{\mathbf{z}}\times\nabla\psi$,
where 
\begin{equation}
\psi=\delta\psi\cos\left(2\pi x/L_{x}\right)\cos\left(\pi y/L_{y}\right).
\end{equation}
The characterizing parameters follow \cite{Daughton2006pp}, and are listed
in Table \ref{tab:Parameters-for-GEM} for convenience. Length,
time, speed, and magnetic field strength are normalized over $d_{i0}$,
$\Omega_{ci0}^{-1}$, and $v_{A0}$, respectively, where $B_{0}$
defines the ion cyclotron frequency $\Omega_{ci0}=eB_{0}/m_{i}$ and
Alfv\'en speed $v_{A0}=B_{0}/\sqrt{\mu_{0}n_{0}m_{i}}$. The resolutions
for five- and ten-moment runs are $4095\times2047$, about $8$ cells
per $d_{e0}$. The PIC run employs a $5760\times2880$ grid ($12$
cells per $d_{e0}$) populated by about $2.7\times10^{9}$ particles
from each species, and resolves Debye length.

Reconnection in the setup described above has been extensively studied
using various numerical models and with different parameters\cite{birn_2001a,Daughton2006pp,Schmitz2006pp,HesseWinske1993grl,Kuznetsova2001jgr,YinWinske2001jgr}.
Particularly, as mentioned in Sec.\ref{sec:intro}, Hesse \textit{et
al.}\cite{HesseWinske1993grl,HesseWinske1994jgr,Kuznetsova2001jgr} and Yin \textit{et al.}\cite{YinWinske2001jgr,YinWinske2002pp,YinWinske2003pp} studied similar systems using modified
hybrid and Hall MHD codes that evolve full $\mathbf{P}_{e}$ with
help of a relaxation term. Our study differs from theirs in the following
important aspects. First, as discussed in Sec.\ref{sec:intro} and
Sec.\ref{sec:closure}, the underlying physical justification of 
their relaxation term is fundamentally different from ours. Second, their 
studies were limited to smaller system sizes. Another interesting study was the
Vlasov simulation by Schmitz and Grauer\cite{Schmitz2006pp},
in which the Vlasov equations are directly evolved, retaining full
kinetic effects. Though their study also used small domain sizes ($25.6d_{i0}\times12.8d_{i0}$),
it agrees well in certain details with our larger targeted and PIC
runs (particularly the latter), as we will show.

\begin{table*}
\begin{tabular}{|c|c|c|c|c|c|c|c|c|c|}
\hline 
Parameter & $L_{x}/d_{i0}$ & $L_{y}/d_{i0}$ & $m_{i}/m_{e}$ & $v_{A0}/c$ & $2\mu_{0}n_{0}k_{B}\left(T_{i0}+T_{e0}\right)/B_{0}^{2}$ & $n_{b}/n_{0}$ & $T_{i0}/T_{e0}$ & $\lambda_{B}/d_{i0}$ & $\delta\psi/B_{0}d_{i0}$\tabularnewline
\hline 
\hline 
Value & $100$ & $50$ & $25$ & $1/15$ & $1$ & $0.3$ & $5$ & $0.913$ & $0.1$\tabularnewline
\hline 
\end{tabular}

\caption{\label{tab:Parameters-for-GEM}Summary of parameters for simulations of the Harris sheet reconnection problem.}
\end{table*}

\subsection{\label{sub:gem-flux}Reconnected flux}

\begin{figure}
\includegraphics[width=1\columnwidth]{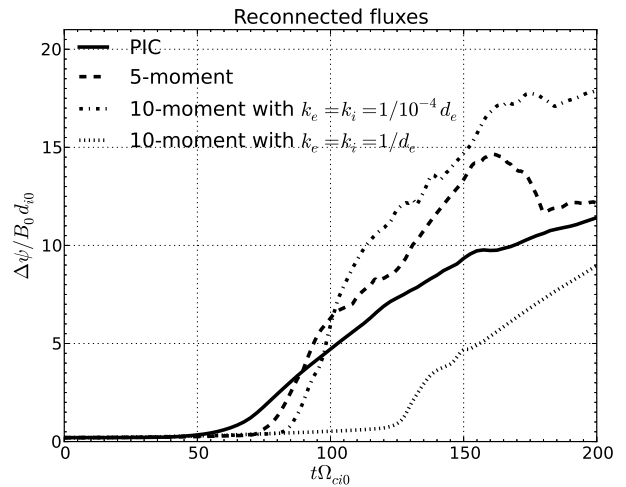}\caption{\label{fig:Reconnected-flux-vs-Time}Reconnected flux $\Delta\psi$
vs. time from simulations of the Harris sheet reconnection problem using different models.}

\end{figure}

The reconnected fluxes, $\Delta\psi$, are presented in Figure (\ref{fig:Reconnected-flux-vs-Time}).
Here, $\Delta\psi$ is calculated by integrating $d\psi=-\left|B_{y}\right|dx/2$
from the left boundary center to the right boundary center, and is
normalized over $B_{0}d_{i0}$. The slope of a $\Delta\psi-t$ curve,
$\partial\Delta\psi/\partial t$, is called the reconnection rate. 

In Fig. \ref{fig:Reconnected-flux-vs-Time}, the different curves do
not agree well with one another when $\Delta\psi$ rises significantly (indicating
onset of reconnection) and and overall slope (indicating the average reconnection
rate). These discrepancies could be caused by the different natures
of PIC and multi-fluid moment models (in terms of certain microinstabilities,
for example), and also by the presence of plasmoids. As will be visualized
in Sec.\ref{sub:gem-uze}, plasmoids are readily generated in the
five-moment run (dashed curve) and the ten-moment runs with $k_{e0}=k_{i0}=1/10^{-4}d_{e0}$
(dashed-dotted curve). Plasmoids accelerate reconnection and lead to early rise
of $\Delta\psi$. In comparison, the targeted run (fine dotted curve) appears to be marginally
stable to plasmoids, where tiny plasmoids appear only very transiently
around $t=150\Omega_{ci0}^{-1}$ and are immediately merged, corresponding
to a much more gentle slope with a tiny ``bump'' around $t=150\Omega_{ci0}^{-1}$.
Finally, the PIC run (solid curve) shows no plasmoid at all and has the most gentle
slope.

Though the physics of plasmoids is not the primary concern in this
study, it modifies time dependences of reconnected fluxes significantly.
Thus, when comparing different runs, we should select frames when
the same amount of fluxes are newly reconnected, i.e., when $\Delta\psi-\Delta\psi\left(t=0\right)$
are the same. In such time frames, different runs might be considered
in a similar stage of the evolution. We will follow this criterion
to select comparable frames in the rest of this study.

\subsection{\label{sub:gem-uze}Structures of electron current layer}

\begin{figure}
\includegraphics[width=0.98\columnwidth]{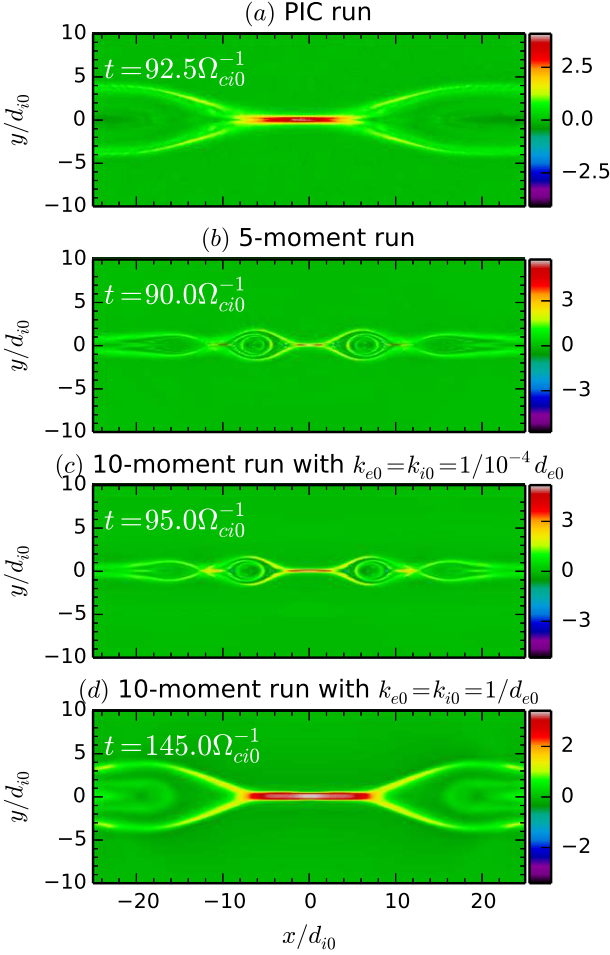}\caption{\label{fig:gem-vze}Color contours of out-of-plane electron velocity,
$u_{z,e}$, in simulations of the Harris sheet reconnection problem using different models.
The frames are selected when $\Delta\psi-\Delta\psi\left(t=0\right)\approx2.5$.}
\end{figure}

Fig. \ref{fig:gem-vze} shows snapshots of electron out-of-plane
velocities, $u_{z,e}$, near the domain centers. These snapshots are
taken at times when $\Delta\psi-\Delta\psi\left(t=0\right)\approx2.5B_{0}d_{i0}$.
It is readily seen that the five-moment run (Panel (b)) and ten-moment
run with $k_{e0}=k_{i0}=1/10^{-4}d_{e0}$ (Panel (c)) generate a chain
of plasmoids, and are remarkably similar to each other. Such similarity
is not surprising, since the ten-moment run with $k_{e0}=k_{i0}=1/10^{-4}d_{e0}$
approaches the five-moment limit. The immediate generation of plasmoid
chains also indicates that, in these two simulations, electron inertia
alone is not sufficient to prevent the electron current layer from
rapidly thinning down to the $d_{e0}$ scale. 

The targeted run (Panel (d)), in comparison, contains no plasmoid
at this time, and is rather similar to the PIC result (Panel (a))
in terms of lengths and thicknesses of the current layers, magnitudes
of the maximum $u_{z,e}$, and opening angles of the separatrices.
It should be pointed out that the frames shown here might not be precisely
timed when $\Delta\psi-\Delta\psi\left(t=0\right)=2.5$ due to limited
frames of data points. Taking into account such possible offsets in
timing, the agreement between the targeted run and the PIC run appears to be reasonably good.

\begin{figure}
\includegraphics[width=0.98\columnwidth]{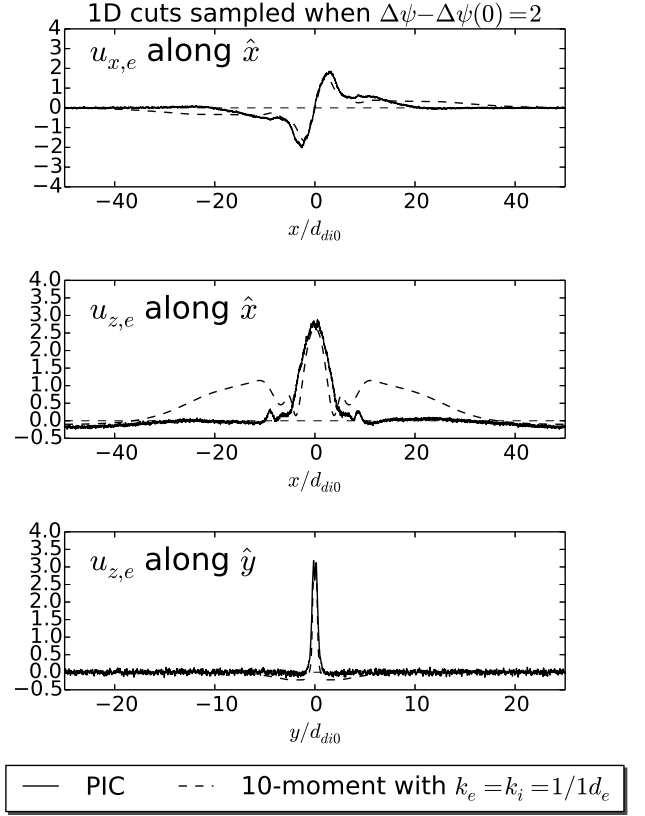}\caption{\label{fig:gem-v-cuts}Cuts of electron velocities in the PIC run
and ten-moment run with $k_{e0}=k_{i0}=1/d_{e0}$ of the Harris sheet
reconnection problem. Top panel is the outflow electron velocities, $u_{x,e}$,
along outflow lines. Middle panel is the out-of-plane electron velocities, $u_{z,e}$,
along outflow lines. Bottom panel is $u_{z,e}$ along inflow lines. Data are at sampled
time when $\Delta\psi-\Delta\psi\left(t=0\right)\approx2$. Note that
the PIC data are slightly shifted along the outflow line to place the stagnation
point right at the domain center.}
\end{figure}

\begin{figure}
\includegraphics[width=0.98\columnwidth]{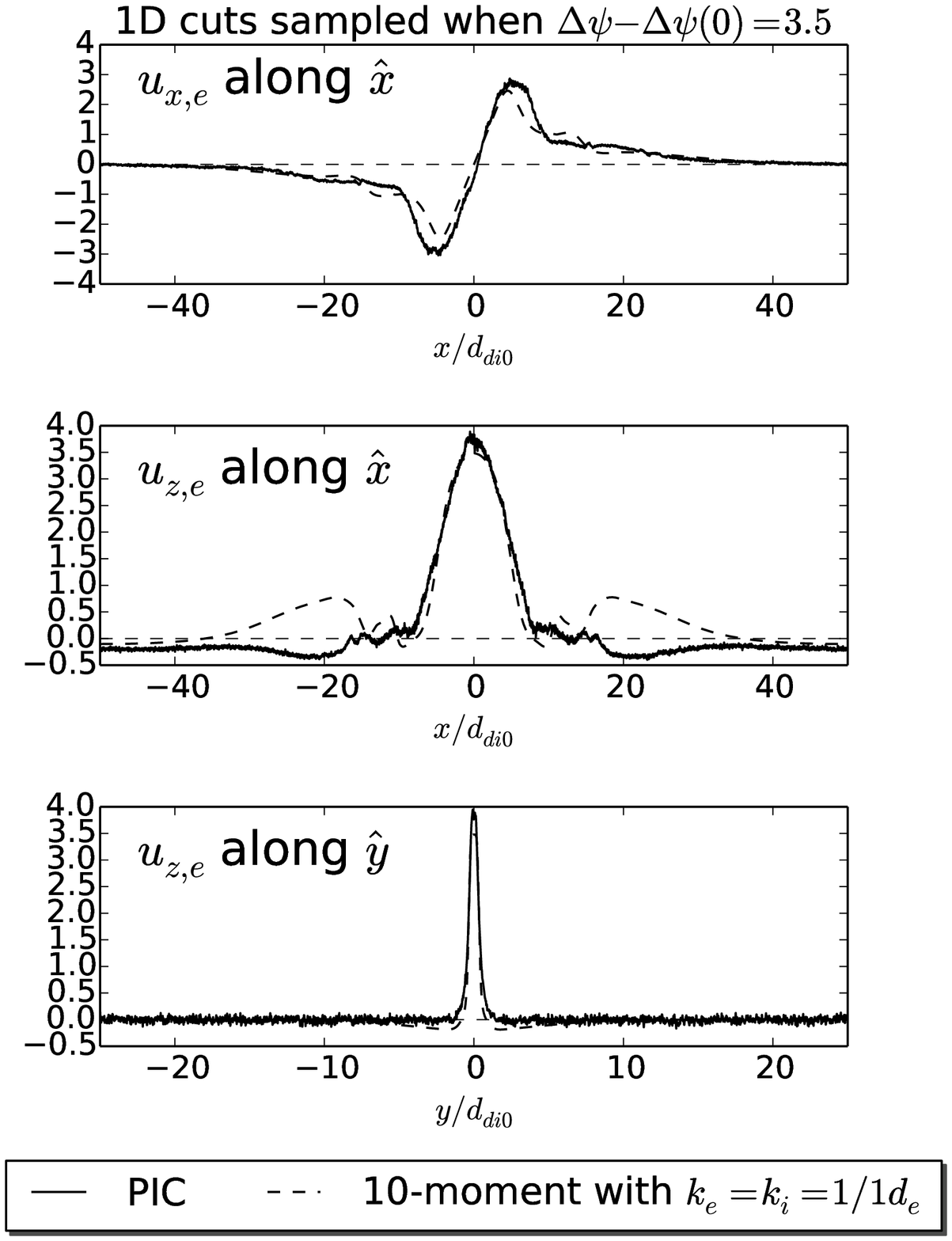}\caption{\label{fig:gem-v-cuts-later}Repeatition of Fig. \ref{fig:gem-v-cuts}
when $\Delta\psi-\Delta\psi\left(t=0\right)\approx3.75$.}
\end{figure}

The agreement between the PIC run and the targeted run can be further
confirmed by Fig. \ref{fig:gem-v-cuts} that shows cuts of electron
flows velocities along inflow and outflow lines from the targeted
run (dashed curves) and the PIC run (solid curves) earlier when $\Delta\psi-\Delta\psi\left(t=0\right)=2$.
Top and middle panels show outflow electron velocities, $u_{x,e}$, and
out-of-plane electron velocities, $u_{z,e}$, along outflow direction.
The two runs agree very well near the central X-point, but the magnitudes
in the PIC run fall faster in the further downstream ``lobe'' regions.
Bottom panel shows $u_{z,e}$ along inflow lines, where the two runs
are remarkably similar. This set of comparisons is repeated in later
time frames when $\Delta\psi-\Delta\psi\left(t=0\right)=3.75$, shown
in Fig. \ref{fig:gem-v-cuts-later}. Similar agreement is observed
again in Fig. \ref{fig:gem-v-cuts-later}.

The excellent agreement between the targeted and PIC runs near the
X-point indicates that the choice of $k_{e0}=k_{i0}=1/d_{e0}$ indeed
correctly handles the dominating physical length scales near the X-point.
The discrepancies in the further downstream lobe regions imply that
$k_{e0}=k_{i0}=1/d_{e0}$ is less appropriate in these regions, since
the characteristic length scales grow larger, as indicated by the
opening of separatrices. Consequently, $k_{e0}$ and $k_{i0}$ should
be reduced accordingly in those regions\@.

\subsection{\label{sub:gem-pe}Relaxation of pressure tensor in the ten-moment
model}

Next, we compare the full pressure tensors from PIC and ten-moment
simulations. This side-by-side comparison directly evaluates the capability
of the ten-moment model to organize the non-gyrotropic, anisotropic
pressure tensors. We should also keep in mind that calculating higher
order moments like the pressure tensor in a PIC simulation is not trivial,
since the data are noisy in nature due to fluctuations of discrete
computational particles.

\begin{figure*}
\includegraphics[width=1.98\columnwidth]{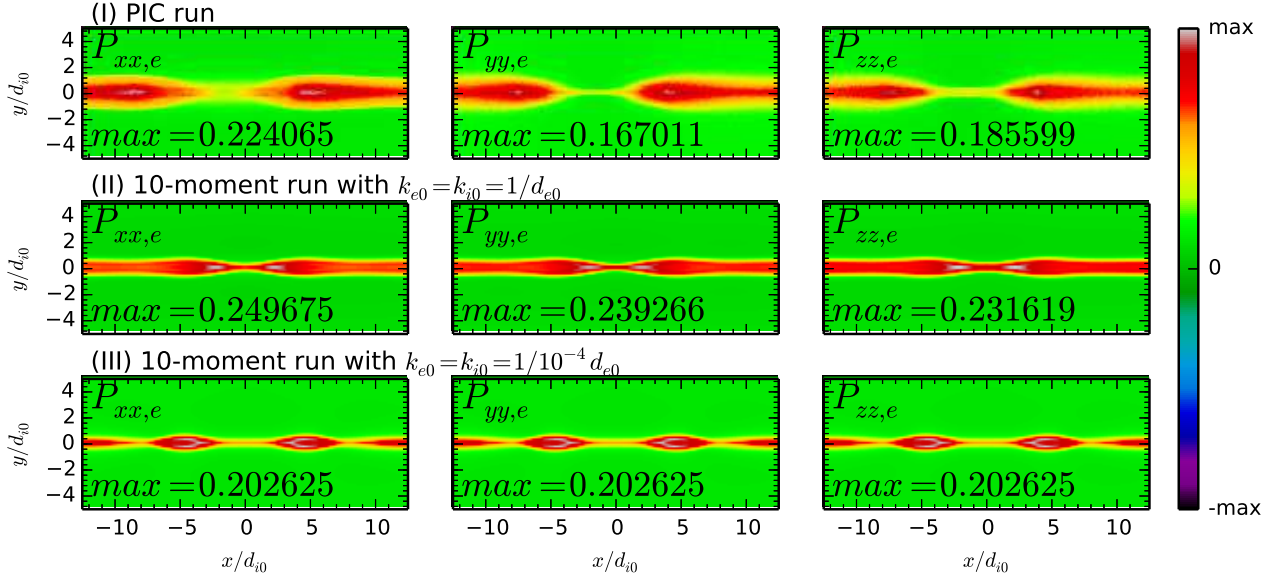}\caption{\label{fig:gem-pe-diag}Color contours of diagonal elements of $\mathbf{P}_{e}$ 
sampled when $\Delta\psi-\Delta\psi\left(t=0\right)=1$.
The three rows (top to bottom) are from the PIC run, the targeted
run, and the ten-moment run with $k_{e0}=k_{i0}=1/10^{-4}d_{e0}$,
respectively. The three columns (left to right) correspond to $P_{xx,e}$,
$P_{yy,e}$, and $P_{zz,e}$, respectively.}
\end{figure*}

Fig. \ref{fig:gem-pe-diag} shows diagonal elements of $\mathbf{P}_{e}$
from the PIC run (first row), the targeted run (second row), and the
ten-moment run with $k_{e0}=k_{i0}=1/10^{-4}d_{e0}$ (bottom row).
The data are sampled early in the quasi-steady states when $\Delta\psi-\Delta\psi\left(t=0\right)=1$.
The global structures of different terms are in agreement between
the PIC and targeted run. The maximum value of $P_{xx,e}$, $P_{yy,e}$,
and $P_{zz,e}$ in the PIC run are $0.224$, $0.167$, and $0.186$
in units of $B_{0}^{2}/\mu_{0}$, obeying the relation $P_{xx,e,max}>P_{zz,e,max}\gtrsim P_{yy,e,max}$.
This relation was also found by previous Vlasov simulation (Fig.(5)
of \refcite{Schmitz2006pp}) on a smaller domain. The corresponding
maximums are $0.25$, $0.239$, and $0.232$ in the targeted run,
obeying a slightly different relation $P_{xx,e,max}\gtrsim P_{zz,e,max}\approx P_{yy,e,max}$.
In other word, the targeted run generates slightly more isotropy than
fully kinetic simulations, which might be improved by a better collisionless
closure in the full 3D regime. In comparison, the three diagonal elements
are almost identical in the ten-moment run with $k_{e0}=k_{i0}=1/10^{-4}d_{e0}$,
since it approaches the isotropic five-moment limit. 

\begin{figure*}
\includegraphics[width=1.98\columnwidth]{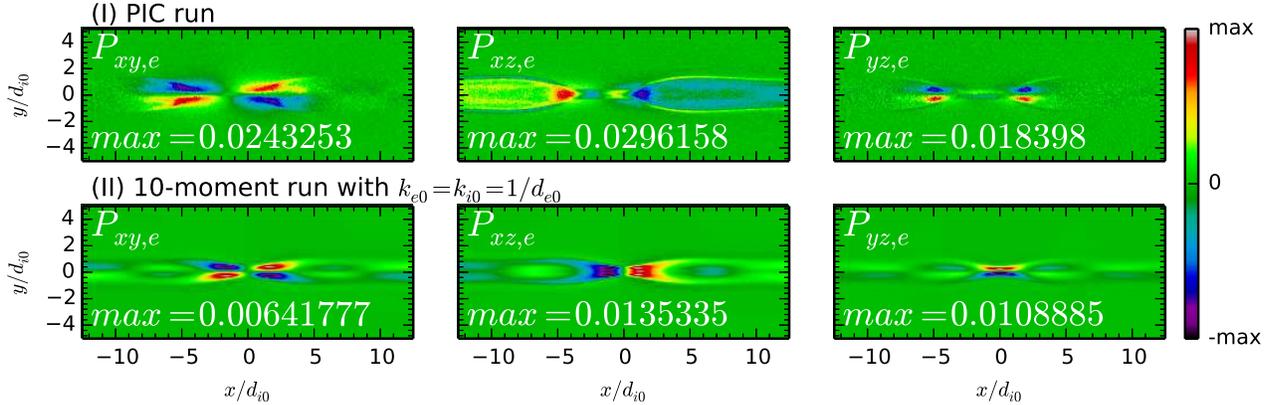}

\caption{\label{fig:gem-pe-off-diag}Color contours of diagonal elements of $\mathbf{P}_{e}$ 
sampled when $\Delta\psi-\Delta\psi\left(t=0\right)=1$.
The top row is from the PIC run, and the bottom row is from the targeted
run. The three columns (left to right) correspond to $P_{xy,e}$,
$P_{yz,e}$, and $P_{xz,e}$, respectively.}
\end{figure*}

Fig. \ref{fig:gem-pe-off-diag} shows off-diagonal elements of $\mathbf{P}_{e}$
from the PIC run (top row) and targeted run (bottom row). The results
from the ten-moment run with $k_{e0}=k_{i0}=1/10^{-4}d_{e0}$ are
not shown here because the off-diagonal elements vanish (almost to
level of machine error) as a result of strong gyrotropization. As
shown in Fig. \ref{fig:gem-pe-off-diag}, the overall structures,
particularly the polarities of these off-diagonal elements in the
PIC run are very well recovered in the targeted run. However, the
relations between magnitudes of terms in different locations do not
agree well. For the quadrupole-shaped $P_{xy,e}$ (first column),
the maximum magnitude in the PIC run ($\sim0.024$) is four times
that in the targeted run ($\sim0.006$). For $P_{xz,e}$ (second column),
the magnitudes near the X-point are close ($\sim0.014$); but in the
further downstream lobe regions, the magnitudes grow to $\sim0.03$
in the PIC run, but decay to much smaller values in the targeted
run run in the same regions. For $P_{yz,e}$ (third column), the magnitudes
near the X-point are $\sim0.005$ in PIC run, smaller than that in
targeted run ($\sim0.01$); the magnitudes grow in the lobe regions
in PIC run to $\sim0.016$, but decay to very small values in Targed.
On the other hand, the small domain Vlasov simulation (Fig.(6) of \refcite{Schmitz2006pp}),
again, showed structures and magnitude relations of these off-diagonal
elements very similar to our PIC run.

From the above comparisons, it is clear that the ten-moment run with
$k_{e0}=k_{i0}=1/10^{-4}d_{e0}$ indeed relaxes the full $\mathbf{P}_{e}$
to a scalar, which is essentially the five-moment limit. It is also
clear that the targeted ten-moment run with $k_{e0}=k_{i0}=1/d_{e0}$
recovers well the full tensor $\mathbf{P}_{e}$ around the X-point.
It does less well in the further downstream lobe regions, though.
On the other hand, though limited by small domain size, previous fully
kinetic Vlasov simulation agrees with the PIC run and the targeted
run in overal structures of various terms, and also demonstrates certain
quantitative relations that are close to our PIC run.

\subsection{\label{sub:gem-ohm}Decomposition of the Ohm's law}

The capability of the ten-moment model to evolve full $\mathbf{P}_{e}$
can also be investigated by decomposing the following form of generalized
Ohm's law (essentially the electron momentum equation) around the
reconnection site:
\begin{eqnarray}
E_{z} & = & -u_{x,e}B_{y}+u_{y,e}B_{x}-\frac{1}{n_{e}\left|e\right|}\left(\frac{\partial P_{xz,e}}{\partial x}+\frac{\partial P_{yz,e}}{\partial y}\right)\nonumber \\
 &  & -\frac{1}{n_{e}\left|e\right|}\left[\frac{\partial}{\partial t}\left(m_{e}n_{e}u_{z,e}\right)+\frac{\partial}{\partial x}\left(m_{e}n_{e}u_{x,e}u_{z,e}\right)\right.\nonumber \\
 &  & \left.+\frac{\partial}{\partial y}\left(m_{e}n_{e}u_{y,e}u_{z,e}\right)\right].\label{eq:ohm-electron-z-10m}
\end{eqnarray}
In a 2D setup, the reconnection rate can be measured by $\left|E_{z}\right|\approx\partial\psi/\partial t$
at the reconnection site, which is an important diagnostic\cite{Vasyliunas1975review}. Historically, PIC and traditional
fluid simulations of 2D anti-parallel reconnection showed vast differences
in the sources of $E_{z}$ at the X-point. PIC simulations showed
that $E_{z}$ is largely supported by the divergence of the off-diagonal
elements of\textbf{ $\mathbf{P}_{e}$}, i.e., $E_{z}^{\text{ng}}\equiv-\left(\partial_{x}P_{xze}+\partial_{y}P_{yze}\right)/n_{e}\left|e\right|$,
while traditional fluid models only permit a scalar pressure, $p_{e}$,
which does not contribute to $E_{z}^{\text{ng}}$. It is thus interesting
to find the contribution of $E_{z}^{ng}$ to $E_{z}$ in ten-moment
simulations to see if it is consistent with the PIC results.

\begin{figure*}
\includegraphics[width=1.98\columnwidth]{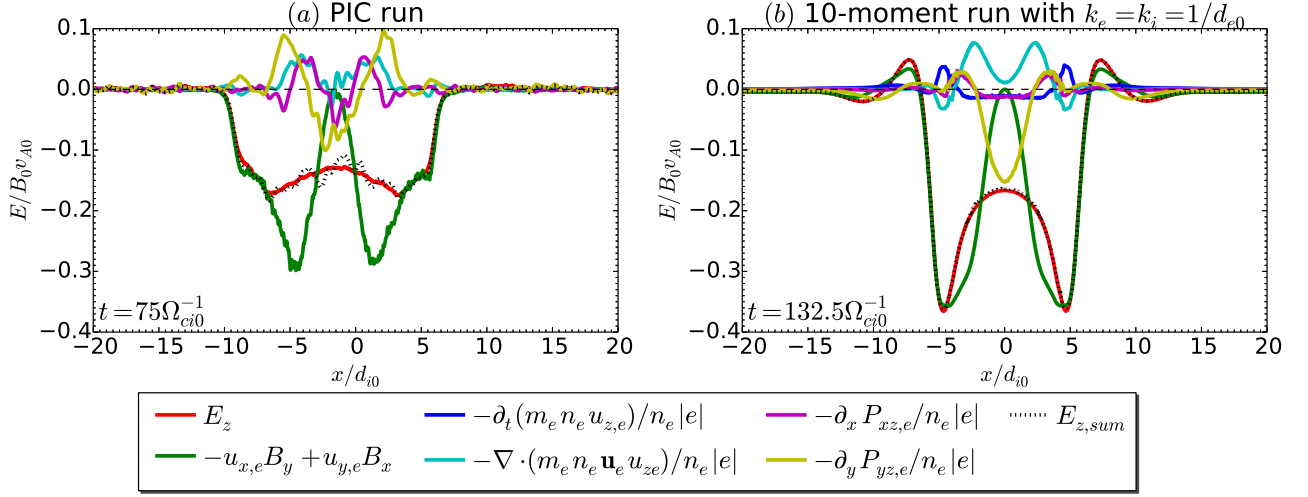}\caption{\label{fig:generalized-ohm-law-decompoistion}Terms consistituting
the generalized Ohm's law, Eq.(\ref{eq:ohm-electron-z-10m}), along
the outflow line (i.e., $x$-axis) from the PIC (left panel) and the
targeted run (right panel). The sampling time is when $\Delta\psi-\Delta\psi\left(t=0\right)=2$.}

\end{figure*}

Fig. \ref{fig:generalized-ohm-law-decompoistion} shows constituting
terms of Eq.(\ref{eq:ohm-electron-z-10m}) along the outflow line
in the PIC run (panel (a)) versus the targeted run (panel (b)). The
sampling time is when $\Delta\psi-\Delta\psi\left(t=0\right)=2$.
Red curves are $E_{z}$, and black dashed curves are sums of terms
on the right-hand-side of equation (\ref{eq:ohm-electron-z-10m}),
denoted by $E_{z,sum}$. In both runs, the magnitudes of $E_{z}$
are comparable, and are both dominated by $-\partial P_{yz,e}/\partial y$
(yellow-green curves) at the stagnation point where $E^{\text{convective}}\equiv-u_{x,e}B_{y}+u_{y,e}B_{x}=0$
(i.e., where green curves touch zero). This agreement cofirms that
ten-moment run is evolving $\mathbf{P}_{e}$ qualitatively correctly.
The structures of terms in Eq.(\ref{eq:ohm-electron-z-10m}) in the
two runs are also similar. For example, in both runs, $E_{z}^{\text{ng}}$
overshoots in the further downstream regions with comparable magnitudes,
while $-\partial P_{yz,e}/\partial y$ and $E^{\text{convective}}$
overshoot in opposite ways in the downstream
regions.

A major difference is that, in the PIC run, the overshooting of $E_{z}^{\text{convective}}$
is largely cancelled by $-\partial P_{yz,e}/\partial y$, and the
resulting $E_{z}$ is relatively flat, while in the targeted run,
little cancellation was shown between the two terms because their
peaks do not overlap. As a result, $E_{z}$ is dominated by $E_{z}^{\text{convective}}$
and in addition, shows overshoots in the downstream regions. 

It should be emphasized that comparison of the various terms in 
the generalized Ohm's law from different types of simulations is 
sensitive to the choice of when in time such comparisons are made. 
The inherent noises in the PIC simulation also make the measurement of
various mean quantities subject to fluctuations. Considering these possible sources
of errors, it is noteworthy that the generalized Ohm's law in the
ten-moment run can achieve rather good agreement with the PIC run,
even with a simple closure (constant $k_{e0}=k_{i0}=1/d_{e0}$
in \eqr{\ref{eq:local-q-closure}}).

\section{\label{sec:Conclusions}Conclusions and Future Work}

We have investigated the multi-fluid moment model in the context of collisionless
magnetic reconnection. Two limits of this model were discussed: the
five-moment limit that evolves scalar pressures, and the ten-moment
limit that evolves full pressure tensors. First, through the example
of the five-moment model, we demonstrated how effects critical
in collisionless reconnection are self-consistently embedded, and
how the resulting five-moment equations formally approach the more widely used Hall
MHD equations under the limits of vanishing electron inertia, infinite speed of light, and quasi-neutrality.
Then, we investigate the capability of the ten-moment
model to evolve full pressure tensors. In order to
approach fully kinetic models, we implemented a local linear \textit{collisionless}
closure in the form of \eqr{\ref{eq:local-q-closure}} to approximate
(divergence of) the heat flux. This simple closure takes a characteristic
wave-number, $k_{0}$, that is constant throughout the simulation
domain to prescribe the scale at which collisionless Landau damping
effectively occurs. When applied to Harris sheet reconnection, the
ten-moment run using this crude closure and $k_{e}=k_{i}=1/d_{e0}$
(which we call the targeted run) yields elements of $\mathbf{P}_{e}$
that are consistent with the PIC results in the vicinity of the reconnection
site, and reproduces electron flows that are remarkably similar to
the PIC ones. In the further downstream lobe regions, however, the
agreement is less satisfactory. In addition, near the X-point, subtle,
but noticeable differences in magnitude relations between different
elements of $\mathbf{P}_{e}$ are also observed.

The discrepancies between the targeted ten-moment run and the PIC run
indicate the need to determine closure parameters, $k_{e0}$ and $k_{i0}$,
from local properties (e.g., scales of local gradients), and the need
for a better closure in the full 3D regime (instead of using the same
$k_{e0}$ and $k_{i0}$ for every direction). As a step to improve
the closure to suit such needs, analytic generalization of \eqr{\ref{eq:local-q-closure}}
is necessary, which will require non-trivial mathematical manipulations.
Methods mentioned in Sec.\ref{sec:closure} to perform non-local
integration through Hilbert transform of the temperature may also
be benchmarked. On the other hand, in order to better understand properties
of the heat flux, it is necessary to compare the approximated heat
flux due to closures in ten-moment simulations with the real heat
flux from fully kinetic simulations, and possibly with heat flux calculated
in a twenty-moment model (not implemented yet) which evolves full
(fourth-order) heat flux tensor. The different ways summarized above
to improve the collisionless closure are left to future work.

The ten-moment model is very attractive for global modeling of large
scale collisionless systems like the Earth's and other planetary magnetospheres.
First, the ten-moment model avoids two well-known deficiencies of
popular MHD-based magnetospheric codes, namely, the lack of full electron
pressure tensor, and the difficulty in effeciently incorporating the
Hall term. In the ten-moment model, both terms (together with other
terms like electron inertia) are self-consistently embedded,
and can be easily implemented using a locally implicit algorithm that
eliminates time step restriction due to quadratic dispersive modes.
On the other hand, the multi-fluid moment model can easily handle
multiple species, which facilitates modeling of situations where
multi-ion-species plays a significant role. In future work, the ten-moment
moment model can be either directly implemented as a base model for
global simulations, or be integrated in an existing global code to
capture necessary physics in localized regions.

To conclude, the multi-fluid moment model provides an alternative
approach to include non-ideal, but collisionless effects in a continuum
code. With a seemingly crude, but physically fundamental closure in the form of \eqr{\ref{eq:local-q-closure}},
the ten-moment model can evolve $\mathbf{P}_{e}$ largely correct when
appropriate closure parameters are chosen. In
future work, different approaches to improve the collisionless closure
should be explored, towards a fully 3D, non-local closure. Direct
application of the ten-moment model to magnetosphere should also
be sought, and could be a crucial step towards integration of kinetic
effects in global magnetospheric models. 

\section*{\label{sec:acknowledgment}Acknowledgment}
This research is supported by 
the NSF-NASA Collaborative Research on Space Weather Grant No. AGS-1338944,
and the U. S. Department of Energy Contract DE-AC02-09CH11466, 
through the Max-Planck/Princeton Center for Plasma Physics and the Princeton Plasma Physics Laboratory. 
Computations were performed on Trillian, a Cray XE6m-200 supercomputer at UNH supported by the NSF MRI program under grant PHY-1229408,
and on facilities at Research Computing Center of the Princeton Plasma Physics Laboratory.
\bibliography{wang-hakim-recon}

\end{document}